\documentclass{elsart}

\usepackage{graphicx}
\usepackage{amssymb}
\usepackage{color}

\begin{document}

\begin{frontmatter}



\title{Asymptotic Mean Time To Failure and Higher Moments for Large, Recursive Networks}


\author{Christian Tanguy}

\address{Orange Labs, CORE/MCN/OTT, 38--40 rue du G\'{e}n\'{e}ral Leclerc, 92794 Issy-les-Moulineaux Cedex 9, France}
\ead{christian.tanguy@orange-ftgroup.com}

\begin{abstract}
This paper deals with asymptotic expressions of the Mean Time To Failure (MTTF) and higher moments for large, recursive, and non-repairable systems in the context of two-terminal reliability. Our aim is to extend the well-known results of the series and parallel cases. We first consider several exactly solvable configurations of identical components with exponential failure-time distribution functions to illustrate different (logarithmic or power-law) behaviors as the size of the system, indexed by an integer $n$, increases. The general case is then addressed: it provides a simple interpretation of the origin of the power-law exponent and an efficient asymptotic expression for the total reliability of large, recursive systems. Finally, we assess the influence of the non-exponential character of the component reliability on the $n$-dependence of the MTTF.
% Text of abstract
\end{abstract}

\end{frontmatter}

% main text

%%%%%%%%%%%%%%%%%%%%%%%%%%%%%%%%%%%%%%%%%%%%%%%%%%%%%%%%%%%%%%%%%%%%%%%%%%%%%%%%%%%%%%%%%%%%%%%%%%
%%%%%%%%%%%%%%%%%%%%%%%%%%%%%%%%%%%%%%%%%%%%%%%%%%%%%%%%%%%%%%%%%%%%%%%%%%%%%%%%%%%%%%%%%%%%%%%%%%
\section{Introduction}
\label{Introduction}

In non-repairable systems, the Mean Time To Failure (MTTF), i.e., the average value $\langle \, t \, \rangle$ of the occurrence of failure is a key parameter of the corresponding reliability \cite{Shooman68,Singh77,KuoZuo}. Analytical expressions of the MTTF have long been well known for simple configurations such as series, parallel, and $k$-out-of-$n$ systems, where each element has a reliability described by the exponential distribution $p(t) = \exp(- \lambda \, t)$ --- the asymptotic dependence of the MTTF for a total number $n$ of elements is $1/n$ (series) and $\ln n$ (parallel) --- or by more complex time distributions \cite{Shooman68,KuoZuo}.

In this work, we show that these results can be extended to recursive, meshed systems of arbitrary size, the latter being indexed by an integer $n$. Our paper is organized as follows. In Section \ref{Simple results}, we briefly survey definitions and simple results that apply for series, parallel and $k$-out-of-$n$ systems. General expressions for the higher moments $\mu_m = \langle \, t^m \, \rangle$ and for the associated cumulants $\kappa_m$ are also given for $k$-out-of-$n$ systems. Section~\ref{Simple meshed architectures} is devoted to a few exactly solvable configurations \cite{Tanguy06a,Tanguy06b,Tanguy06c,TanguyINOC07} of identical components with exponential failure-time distribution functions, which give rise to different behaviors from those of the series-parallels ones as $n$ goes to infinity. These examples allow us to derive the asymptotic expansion of the MTTF when $n$ is large, and pave the way to the general case addressed in Section~\ref{General case}, which can be roughly divided into ``series-like'' and ``parallel-like'' configurations. Section~\ref{Approximate reliability of large recursive systems} provides a simple, approximate expression for the corresponding total reliability and contributes to an understanding of the behavior of large systems, which are an important issue \cite{Kolowrocki04}. We finally investigate in Section~\ref{Nonexponential distribution functions} how the asymptotic $n$-dependence of the MTTF and higher moments is modified by taking non-exponential failure-time distribution functions into account.

\section{Results for simple systems}
\label{Simple results}

\subsection{Definitions}
\label{Definitions}

As explained in many textbooks \cite{Shooman68,Singh77,KuoZuo}), if $R(t)$ is the system's reliability, the MTTF is defined by
\begin{equation}
{\rm MTTF} = \langle \, t \, \rangle = \int_0^{\infty} \, t \, (-R'(t) \, dt) = \int_0^{\infty} \, dt \, R(t) \, ;
\label{MTTF general}
\end{equation}
The higher moments $\mu_m$ are
\begin{equation}
\mu_m = \langle \, t^m \, \rangle = \int_0^{\infty} \, t^m \, (-R'(t) \, dt) = m \, \int_0^{\infty} \, dt \, t^{m-1} \, R(t) \, .
\end{equation}
For instance, the standard deviation $\sigma$ is given by $\sigma^2 = \langle \, t^2 \, \rangle - \langle \, t \, \rangle^2$. The moment generating function $f(z)$ is a formal series defined by
\begin{equation}
f(z) = 1+\sum_{k=1}^{\infty} \, \frac{\langle \, t^m \, \rangle}{m!} \, z^m ;
\end{equation}
by construction, $\langle \, t^m \, \rangle = f^{(m)}(0)$ (the $m$th derivative of $f(z)$ taken at $z = 0$). Since $f(0)=1$, the cumulant generating function $g(z)$ is subsequently defined by
\begin{equation}
f(z) = \exp \left( \sum_{k=m}^{\infty} \, \frac{\kappa_m}{m!} \, z^m \right) = \exp g(z) \, ,
\end{equation}
where $\kappa_m$ is the cumulant of order $m$; $\kappa_2 = \mu_2 - \mu_1^2$ is the variance, $\kappa_3 = \mu_3 - 3 \, \mu_1 \, \mu_2 + 2 \, \mu_1^3$, and so on. Since $g(z) = \ln \langle \, e^{z\, t} \, \rangle$, $\kappa_m = g^{(m)}(0)$.
Using the same integration by parts as in eq.~(\ref{MTTF general}) leads to
\begin{equation}
f(z) = \langle \, e^{z\, t} \, \rangle = 1 + z \, \int_0^{\infty} \, dt \,  e^{z\, t} \, R(t) = 1 + z \, \widetilde{R}(-z) \, ,
\end{equation}
where $\widetilde{R}$ is the Laplace transform of the reliability.

Calculations are often performed by replacing $R(t)$ with an exponential function, $R_{\rm exp}(t) = \exp (- \lambda \, t)$. We have then
\begin{equation}
\langle \, t^m \, \rangle^{\rm (exp)} = m \, \int_0^{\infty} \, dt \,  e^{- \lambda \, t} \, t^{m-1} =  \frac{m!}{\lambda^m} \,
\label{mu_m exponentielle}
\end{equation}
and $\displaystyle \langle \, e^{z\, t} \, \rangle = \left( 1 - \frac{z}{\lambda} \right)^{-1}$. Consequently,
\begin{equation}
g^{\rm (exp)}(z) = - \ln \left( 1 - \frac{z}{\lambda} \right) \, ,
\end{equation}
from which
\begin{equation}
\kappa^{\rm (exp)}_m = \frac{(m-1)!}{\lambda^m} \, .
\label{kappa_m exponentielle}
\end{equation}

When $p(t) = \exp(- \lambda \, t)$ is the common reliability of the system's elements, the total reliability is
\begin{equation}
R(t) = {\mathcal R}(p(t)) = {\mathcal R}\left( e^{- \lambda \, t} \right) \, ,
\end{equation}
where ${\mathcal R}(p)$ is the reliability polynomial; using the change of variable $\displaystyle t = - \frac{\ln p}{\lambda}$, we get
\begin{eqnarray}
{\rm MTTF} & = & \int_0^{\infty} \, dt \, R(t) = \frac{1}{\lambda} \, \int_0^{\infty} \, d(\lambda \, t) \, {\mathcal R}\left( e^{- \lambda \, t} \right) \, \nonumber \\
& = &  \frac{1}{\lambda} \, \int_0^{1} \, \frac{dp}{p} \, {\mathcal R}(p) \, ,
\label{integrale de Rel sur p}
\end{eqnarray}
and
\begin{equation}
\langle \, t^m \, \rangle = \frac{m}{\lambda^m} \, \int_0^{1} \, \frac{dp}{p} \, (- \ln p)^{m-1} \, {\mathcal R}(p) \, .
\label{mu_m general pour distribution exponentielle}
\end{equation}
We make full use of eqs.~(\ref{integrale de Rel sur p}) and (\ref{mu_m general pour distribution exponentielle}) in the next sections. The moment generating function also reads
\begin{eqnarray}
\langle \, e^{z\, t} \, \rangle & = & 1+ z \, \int_0^{\infty} \, dt \, e^{z\, t} \, {\mathcal R}\left( e^{- \lambda \, t} \right)  \, \nonumber \\
& = &  1+ \frac{z}{\lambda} \, \int_0^{1} \, \frac{dp}{p^{1+z/\lambda}} \, {\mathcal R}(p) \, .
\label{moyenne exp z t}
\end{eqnarray}

\subsection{A few results}

\begin{itemize}
\item For $n$ elements in series, ${\mathcal R}_n(p) = p^n$, so that $R_n(t) = \exp (- n \, \lambda \, t)$, and $\displaystyle {\rm MTTF}^{(\rm series)}_{n} = \frac{1}{n \, \lambda}$ \cite{Shooman68,KuoZuo}. Equations(~\ref{mu_m exponentielle}) and (~\ref{kappa_m exponentielle}) become
\begin{eqnarray}
\mu^{\rm (series)}_m & = & \frac{m!}{n^m \, \lambda^m} \\
\kappa^{\rm (series)}_m & = & \frac{(m-1)!}{n^m \, \lambda^m} \, .
\end{eqnarray}

\item If the $n$ elements are in parallel, then $R_n(t) = 1 - (1 - \exp (- \lambda \, t))^n$, and \cite{Shooman68,KuoZuo}
\begin{equation}
{\rm MTTF}^{(\rm parallel)}_{n} = \frac{1}{\lambda} \, \sum_{i = 1}^{n} \, \frac{1}{i} = \frac{1}{\lambda} \, (\psi(n+1) - \psi(1))\rightarrow \frac{1}{\lambda} \, \left( \ln n + \mathbf{C} + \frac{1}{2 \, n} + \cdots \right)
\end{equation}
for $n$ large, where $\displaystyle \psi(z) = \frac{d \ln \Gamma(z)}{dz}$ is the digamma function and $\mathbf{C} \approx 0.577216$ is the Euler gamma constant (see eq.~8.362.1 of \cite{Gradshteyn94}). In this case, the MTTF diverges as $n$ goes to infinity.

\item For $k$-out-of-$n$:G systems, the reliability polynomial is \cite{Shooman68,KuoZuo}
% page 209 du Shooman, c'est r-out-of-n
\begin{equation}
{\mathcal R}_{k,n}(p) = \sum_{i=k}^{n} \, \frac{n!}{(n-i)! \, i!} \, p^{i} \, (1 - p)^{n-i} \, .
\end{equation}
All the moments could be obtained from eq.~(\ref{mu_m general pour distribution exponentielle}). However, the formula for the cumulants being eventually extremely simple, we focus on these parameters. Starting from eq.~(\ref{moyenne exp z t}), we   find
\begin{eqnarray}
\langle \, e^{z\, t} \, \rangle & = & 1+ \frac{z}{\lambda} \, \int_0^{1} \, \frac{dp}{p^{1+z/\lambda}} \, \sum_{i=k}^{n} \, \frac{n!}{(n-i)! \, i!} \, p^{i} \, (1 - p)^{n-i} \nonumber \\
& = & 1 + \frac{z}{\lambda} \, \sum_{i=k}^{n} \,  \frac{n!}{(n-i)! \, i!} \, \frac{\Gamma\left(i - \frac{z}{\lambda}\right) \, \Gamma(n+1-i)}{\Gamma\left(n+1 - \frac{z}{\lambda}\right)} \nonumber \\
& = & 1 + \frac{z}{\lambda} \, \frac{n!}{\Gamma\left(n+1 - \frac{z}{\lambda}\right)} \, \sum_{i=k}^{n} \, \frac{\Gamma\left(i - \frac{z}{\lambda}\right)}{i!} \nonumber \\
& = &  \frac{n!}{\Gamma\left(n+1 - \frac{z}{\lambda}\right)}\, \frac{\Gamma\left(k - \frac{z}{\lambda}\right)}{(k-1)!} \, ;
\end{eqnarray}
the last equality is proven by induction. Consequently,
\begin{equation}
g(z) = \ln \frac{\Gamma\left(k - \frac{z}{\lambda}\right)}{\Gamma(k)} \, \frac{\Gamma(n+1)}{\Gamma\left(n+1 - \frac{z}{\lambda}\right)} \, ,
\end{equation}
whence (see eq.~8.363.8 of \cite{Gradshteyn94})
\begin{eqnarray}
\kappa_m & = & g^{(m)}(0) = \left( \frac{-1}{\lambda} \right)^m \, \left( \psi^{(m-1)}(k) - \psi^{(m-1)}(n+1) \right) \nonumber \\
& = & \frac{(m-1)!}{\lambda^m} \, \sum_{i=k}^{n} \, \frac{1}{i^m} \, ;
\label{cumulant ordre m pour k-out-of-n}
\end{eqnarray}
Equation~(\ref{cumulant ordre m pour k-out-of-n}) generalizes the expressions obtained for the series ($k=n$) and parallel ($k=1$) cases, and a similar expression for the variance \cite{Patel76}. When $m > 1$, $\kappa_m$ is bounded by the finite value $\displaystyle \frac{(m-1)!}{\lambda^m} \, \zeta(m)$, where $\zeta(m)$ is the Zeta function.
\end{itemize}

Let us now investigate more complex systems.

\section{Simple recursive architectures}
\label{Simple meshed architectures}

We consider in this section configurations that are not reducible to series-parallel ones. These systems are studied in the two-terminal reliability context, in which the source and the destination are the large, colored nodes in Figs.~\ref{schema echelle de K4},\ref{schema simple fan},\ref{schema double fan},\ref{schema Street 3xn}, and \ref{coupures d'ordre 2 et 3 pour les Street 3xn}. The size of the system is indexed by an integer $n$, which counts the number of elementary building blocks of the whole structure. All the nodes are assumed perfect (they never fail), while the reliability of all the edges is described by $p(t)= \exp(- \lambda \, t)$ (we may omit in the following the explicit reference to time and note  this reliability $p$, for greater generality). These recursive architectures have been solved recently. Our aim is to show that the associated MTTFs, as well as moments and cumulants of higher order, can be calculated exactly and that their quite distinct asymptotic expansions in $n$ give good approximations to the exact results.

\subsection{$K_4$ ladder}
\label{K_4 ladder}

%figure1
\begin{figure}[htb]
\centering
\includegraphics[scale=0.5]{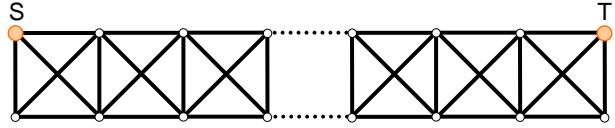}
\caption{$K_4$ ladder architecture.}
\label{schema echelle de K4}
\end{figure}

This architecture, displayed in Fig.~\ref{schema echelle de K4}, is constituted by the repetition of perfect graphs $K_4$ (the fully-connected graph with four nodes). This configuration is exactly solvable even when edges and nodes have distinct reliabilities \cite{Tanguy06c}. For perfect nodes, and with all edge reliabilities equal to $p$, the two-terminal reliability has the simple form
\begin{eqnarray}
{\mathcal R}_{n\geq 1}(p) & = & \alpha_+(p) \, \zeta_+^n(p) + \alpha_-(p) \, \zeta_-^n(p) , \label{Rn analytique echelle K4} \\
\zeta_{\pm}(p) & = & \frac{p}{2} \, \left[ 2 + 4 \, p - 14 \, p^2 +13 \, p^3 - 4 \, p^4 \pm \sqrt{{\mathcal A}(p)} \right] , \label{zeta+ analytique echelle K4} \\
\alpha_{\pm}(p) & = &  \frac{1+p}{4} \pm \frac{2 + 2 \, p +10 \, p^2 -27 \,
p^3 +19 \, p^4-4 \, p^5}{4 \, \sqrt{{\mathcal A}(p)} } , \label{alpha+ analytique echelle K4} \\
{\mathcal A}(p) & = & 4 + 32 \, p^2 -204 \, p^3 +452 \, p^4 - 516 \, p^5 + 329 \, p^6 -112 \, p^7 +16 \, p^8 .
\end{eqnarray}

%figure2
\begin{figure}[htb]
\centering
\includegraphics[scale=0.75]{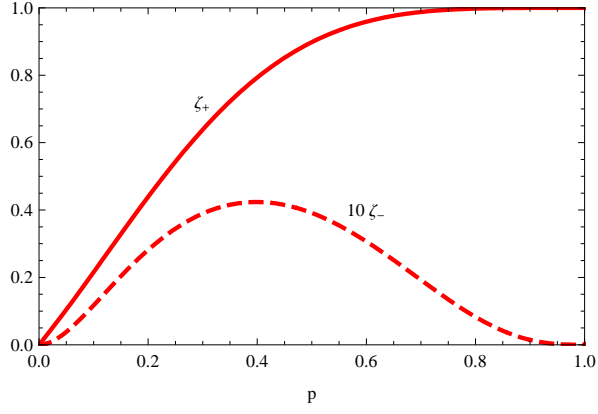}
\caption{$\zeta_{\pm}$ for the $K_4$ ladder.}
\label{zeta plus ou moins echelle de K4}
\end{figure}

\begin{figure}[htb]
\centering
\includegraphics[scale=0.75]{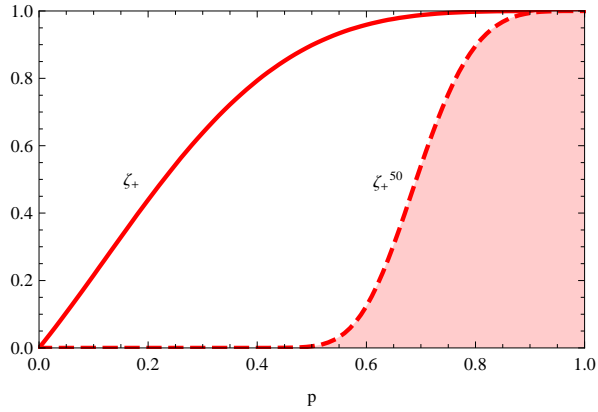}
\caption{Different powers of $\zeta_+$ as a function of $p$.}
\label{puissances de zeta plus echelle de K4}
\end{figure}

The two eigenvalues are displayed in Fig.~\ref{zeta plus ou moins echelle de K4} as functions of $p$. As $n$ increases, the contribution of $\zeta_+$ prevails over that of $\zeta_-$. Therefore, when $n$ is large,
\begin{equation}
{\mathcal R}_{n}(p) \approx \alpha_+(p) \, \zeta_+^n(p) \, ,
\label{approximation Rn par zeta+}
\end{equation}
because the contribution of $\zeta_-^n(p)$ vanishes exponentially. The approximation given by eq.~(\ref{approximation Rn par zeta+}) lies at the heart of our method for deriving the MTTF's asymptotic expansion. Inserting it into eq.~(\ref{integrale de Rel sur p}) gives
\begin{equation}
\langle \,t \, \rangle \approx \frac{1}{\lambda} \, \int_0^{1} \, \frac{dp}{p} \, \alpha_+(p) \, \zeta_+^n(p) \, .
\label{approximation MTTF par alpha+ zeta+ puissance n}
\end{equation}
Here, the lower bound (zero) does not play a significant role because $\zeta_{\pm}(0) = 0$. As $n$ increases, $\zeta_+^n(p)$ is negligible except close to $p = 1$, as illustrated in Fig.~\ref{puissances de zeta plus echelle de K4}.

%figure3

The gist of the asymptotic expansion is therefore to consider $\alpha_+$ and $\zeta_+$ in the vicinity of unity. Setting $q = 1-p$, we have from eqs.~(\ref{zeta+ analytique echelle K4})--(\ref{alpha+ analytique echelle K4})
\begin{eqnarray}
\alpha_+(1-q) & = &  1 - 2 \, q^3 + 4 \, q^5 - 3 \, q^6 + 6 \, q^7 + \cdots \, , \label{DL alphaplus en q}\\
- \ln \zeta_+(1-q) & = &  q^4 + 2 \, q^5 -4 \, q^7 + \frac{9}{2} \, q^8 + \cdots \, . \label{DL ln zetaplus en q}
\end{eqnarray}
Note that $\zeta_+(1) = \alpha_+(1) = 1$ because ${\mathcal R}_{n}(p=1) = 1$. We can write
\begin{equation}
\langle \,t \, \rangle \approx \frac{1}{\lambda} \, \int_0^{1} \, \frac{dq}{1-q} \, \alpha_+(1-q) \, \exp \big[ - n \, (- \ln \zeta_+(1-q))\big]  \, .
\label{approximation MTTF par alpha+ zeta+ puissance n version en q}
\end{equation}
At this point, we have to rescale the variable $q$ in order to extract the asymptotic behavior of the integral. Equation~(\ref{DL ln zetaplus en q}) gives
\begin{equation}
\exp[- n \, (- \ln \zeta_+(1-q))] = \exp[- n (q^4 + 2 \, q^5 -4 \, q^7 + \cdots) ] \, ;
\label{approximation zeta+ puissance n en q 1}
\end{equation}
this suggests setting $\tau = n \, q^4$, or equivalently $q = \tau^{1/4} \, n^{-1/4}$, so that
\begin{equation}
\exp \big[ - n \, (- \ln \zeta_+(1-q))\big] = e^{- \tau} \, \exp\big[- n \, (2 \, \frac{\tau^{5/4}}{n^{5/4}} - 4 \, \frac{\tau^{7/4}}{n^{7/4}} + \cdots) \big] \, .
\label{approximation zeta+ puissance n en q 2}
\end{equation}
In the last exponential, the argument is $-2 \, \tau^{5/4} \, n^{-1/4} + 4 \, \tau^{7/4} \, n^{-3/4} + \cdots$. Because of the $\exp(- \tau)$ factor, we can neglect the contribution of large $\tau$'s, so that when $n$ is large, we only need to expand the second exponential and all other factors in the limit $\tau \to 0$:
\begin{equation}
\langle \,t \, \rangle \approx \frac{1}{\lambda} \, \int_0^{n} \, \frac{1}{4 \, n^{1/4}} \, \frac{d\tau \, \tau^{-3/4}}{1-\frac{\tau^{1/4}}{n^{1/4}}} \, \big(1-2 \, \frac{\tau^{3/4}}{n^{3/4}} + 4 \, \frac{\tau^{5/4}}{n^{5/4}} + \cdots \big) \, e^{- \tau} \, \exp\big[- (2 \, \frac{\tau^{5/4}}{n^{1/4}} - 4 \, \frac{\tau^{7/4}}{n^{3/4}} + \cdots) \big] \, .
\label{approximation zeta+ puissance n en q 3}
\end{equation}

The error made by replacing the upper bound of the integral by $+ \infty$ vanishes exponentially as $n$ goes to infinity. Keeping only the prevailing terms in each factor of eq.~(\ref{approximation zeta+ puissance n en q 3}) leads to
\begin{eqnarray}
\langle \,t \, \rangle & \rightarrow & \frac{1}{\lambda} \, \int_0^{\infty} \, \frac{1}{4 \, n^{1/4}} \, d\tau \, \tau^{-3/4} \,  \, e^{- \tau} \nonumber \\
& = & \frac{1}{\lambda} \, \frac{1}{4 \, n^{1/4}} \, \Gamma(1/4) = \frac{1}{\lambda} \, \frac{\Gamma(5/4)}{n^{1/4}} \approx \frac{0.906402}{\lambda \, n^{1/4}} \, .
\label{leading approximation echelle K4}
\end{eqnarray}
For the leading-order term (and this term only), $\alpha_+$ does not play any role since it may safely be replaced with 1. The asymptotic $n$-dependence is not $1/n$ or $\ln n$ anymore as in the series and parallel cases, but a power-law, $n^{-1/4}$, which slowly decreases with $n$. The following terms of the expansion may be derived easily by expanding all the factors in eq.~(\ref{approximation zeta+ puissance n en q 3}):
\begin{equation}
\lambda \, \langle \,t \, \rangle = \frac{\Gamma(5/4)}{n^{1/4}} + \frac{17}{32} \, \frac{\Gamma(3/4)}{n^{3/4}} - \frac{3}{4 \, n} - \frac{293}{512} \, \frac{\Gamma(5/4)}{n^{5/4}} + \cdots \, .
\label{leading approximation echelle K4 2}
\end{equation}

\begin{figure}[htb]
\centering
\includegraphics[scale=0.75]{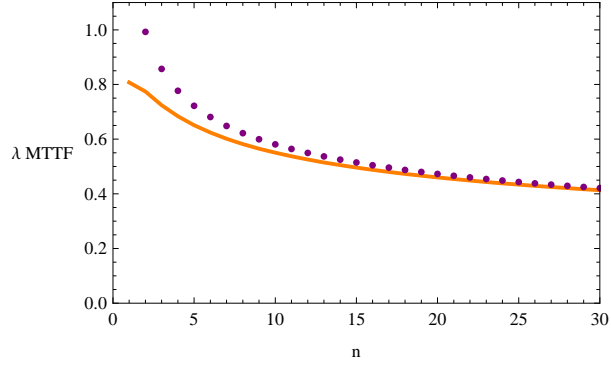}
\caption{Comparison between exact and asymptotic (see eq.~(\ref{leading approximation echelle K4 2})) MTTF.}
\label{Comparaison MTTF exact et asymptotique echelle de K4}
\end{figure}

The exact MTTF is obtained straightforwardly by using eq.~(\ref{integrale de Rel sur p}) and the value ${\mathcal R}_{n}(p)$ deduced from the three-term recursion relation at the origin of eq.~(\ref{Rn analytique echelle K4}):
\begin{eqnarray}
{\mathcal R}_{n}(p) & = & (\zeta_+(p) + \zeta_-(p)) \, {\mathcal R}_{n-1}(p) - \zeta_+(p) \, \zeta_-(p) \, {\mathcal R}_{n-2}(p) \nonumber \\
& = & p \, (2 + 4 \, p - 14 \, p^2 +13 \, p^3 - 4 \, p^4) \, {\mathcal R}_{n-1}(p) \nonumber \\
& & - p^3 \, (4 - 18 \, p + 36 \, p^2 - 42 \, p^3 + 30 \, p^4 - 12 \, p^5 + 2 \, p^6) \, {\mathcal R}_{n-2}(p) \, ,
\label{relation recurrence Rn echelle K4}
\end{eqnarray}
with ${\mathcal R}_{0}(p) = 1$ and ${\mathcal R}_{1}(p) = p \, (1 + 2 \, p - 7 \, p^3 + 7 \, p^4 -2 \, p^5)$. The exact values of the MTTF are then obtained by a simple integration of ${\mathcal R}_{n}(p)$. The exact and the asymptotic (limited to the first three terms of the expansion) results for the MTTF are plotted in Fig.~\ref{Comparaison MTTF exact et asymptotique echelle de K4}. Even for moderate values of $n$, the agreement between the two  is good.

%figure4

Following this method, we also find the asymptotic expansion of $\langle \,t^2 \, \rangle$ by adding the factor $\displaystyle - 2 \, \ln\left(1-\frac{\tau^{1/4}}{n^{1/4}}\right)$ and another $1/\lambda$ in eq.~(\ref{approximation zeta+ puissance n en q 3}). As regards the leading term of the expansion, a mere factor $2 \, \frac{\tau^{1/4}}{n^{1/4}} \, \lambda^{-1}$ is added in the integral. Finally
\begin{equation}
\lambda^2 \, \langle \,t^2 \, \rangle = \frac{\sqrt{\pi}}{2 \, \sqrt{n}} + \frac{17}{12 \, n} - \frac{13 \, \Gamma(9/4)}{10 \, n^{5/4}} - \frac{221 \, \sqrt{\pi}}{480 \, n^{3/2}} + \cdots \, .
\label{leading approximation echelle K4 3}
\end{equation}
Further terms can be routinely obtained using mathematical software such as {\sc Mathematica}.

After simplification, the variance of the distribution is therefore deduced to behave as
\begin{equation}
\langle \,t^2 \, \rangle - \langle \,t \, \rangle^2 = \frac{1}{\lambda^2} \, \left[ \left(\frac{\sqrt{\pi}}{2}- \Gamma(5/4)^2 \right) \, \frac{1}{\sqrt{n}} + \frac{17}{12 \, n} \, \left(1 - \frac{3 \, \pi \, \sqrt{2}}{16} \right) + \cdots \right] \, .
\label{leading approximation variance MTTF echelle K4}
\end{equation}
We could perform similar calculations for higher moments or cumulants, and again would find asymptotic power-law behaviors.

%%%%%%%%%%%%%%%%%%%%%%%%%%%%%%%%%%%%%
\subsection{Generalized fan}
\label{Generalized fan architecture}

%figure5
\begin{figure}[htb]
\centering
\includegraphics[scale=0.5]{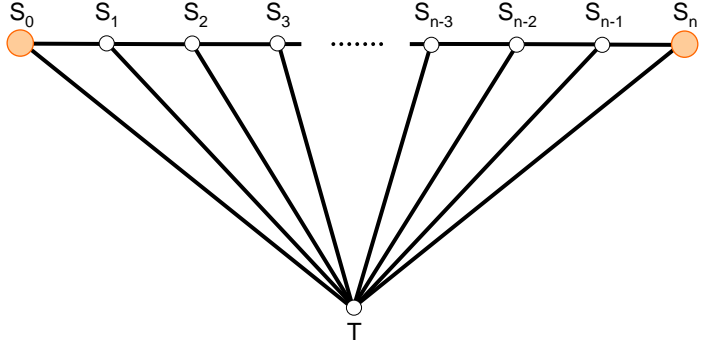}
\caption{Generalized fan: the source is $S_0$, the destination $S_n$.}
\label{schema simple fan}
\end{figure}

This architecture, displayed in Fig.~\ref{schema simple fan}, has been considered in previous studies \cite{Aggarwal75,Neufeld85,Gordon01} and recently solved for the two-terminal reliability ${\mathcal R}_n$ between $S_0$ and $S_n$ \cite{Tanguy06b}. For perfect nodes,
\begin{equation}
{\mathcal R}_n = \frac{p^2}{\big( 1-p \, (1-p)\big)^2} + p^n \, (1-p)^{n+2} \, \left[\frac{n \, p}{\big( 1-p \, (1-p)\big)} + \frac{1+p^2}{\big( 1-p \, (1-p)\big)^2} \right] \, .
\label{Rel2 de n simple fan}
\end{equation}
When $n \to \infty$, ${\mathcal R}_n$ clearly tends to the constant $\displaystyle {\mathcal R}_{\infty} = \frac{p^2}{\big( 1-p \, (1-p)\big)^2} \neq 1$ ($p \, (1-p)$ is always less than 1/4, so the last contribution in eq.~(\ref{Rel2 de n simple fan}) decreases faster than $4^{-n}$). This stems from the existence of one path with a finite number of hops, namely $S_0 \to T \to S_n$. The MTTF's asymptotic behavior is therefore different from that of the preceding section: it does not vary with $n$. This is also true for higher moments, with
\begin{equation}
\langle \,t^m \, \rangle_{\infty} = \frac{m}{\lambda^m} \, \int_0^1 \, dp \, (- \ln p)^{m-1} \, \frac{p}{\big( 1-p \, (1-p)\big)^2} \, ,
\label{moment ordre m simple fan general}
\end{equation}
where $\langle \,t^m \, \rangle_{\infty} = \lim_{n \to \infty} \, \langle \, t^m \, \rangle_n$. The first of these integrals are
\begin{eqnarray}
\lambda \, \langle \, t \, \rangle_{\infty} & = & \frac{9 + 2 \, \pi \, \sqrt{3}}{27} \approx 0.736400 \, ,\\
\lambda^2 \, \langle \, t^2 \, \rangle_{\infty} & = & \frac{2}{9} \, \psi'(1/3) - \frac{4}{27} \, \pi^2  \approx 0.781302 \, ,
\label{premiers moments simple fan}
\end{eqnarray}
where $\psi'$ is the derivative of the digamma function $\psi$. From the first values of $\langle \, t^m \, \rangle$, we can infer the general result
\begin{eqnarray}
\lambda^m \, \langle \, t^m \, \rangle_{\infty} & = & (-1)^m \, \frac{m}{3^{m+1}} \, \left( 1 + \frac{1}{2^{m-1}} \right) \, (\psi^{(m-1)}(1/3)- \psi^{(m-1)}(2/3)) \nonumber \\
& & - \frac{m!}{3^{m-1}} \, \left( 1 - \frac{1}{2^{m-2}} \right) \, (3^{m-2}-1) \, \zeta(m-1) \, .
\label{moments generaux simple fan}
\end{eqnarray}
Depending on the parity of $m$, the difference $\psi^{(m-1)}(1/3)- \psi^{(m-1)}(2/3)$ may actually be further simplified (leaving only $\psi^{(m-1)}(1/3)$ for $m$ even, or powers of $\pi$ for $m$ odd). It is easy to prove that, asymptotically,
\begin{equation}
\langle \, t^m \, \rangle_{\infty} \sim \frac{m!}{2^m \, \lambda^m} \, .
\end{equation}
In that limit, it looks as if only the $S_0 \to T \to S_n$ connection exists.

%%%%%%%%%%%%%%%%%%%%%%%%%%%%%%%%%%%%%
\subsection{Double fan}
\label{Double fan}

%figure6
\begin{figure}[htb]
\centering
\includegraphics[scale=0.5]{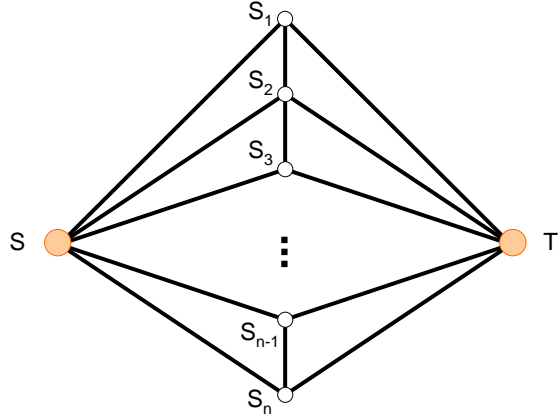}
\caption{Double fan: the source is $S$, the destination $T$.}
\label{schema double fan}
\end{figure}

This configuration, displayed in Fig.~\ref{schema double fan}, is a slight generalization of $n$ double links in parallel. As $n \to \infty$, there is an infinity of paths of finite length connecting $S$ to $T$; for perfect nodes, the associated two-terminal reliability ${\mathcal R}_n$ is \cite{TanguyINOC07}
\begin{equation}
{\mathcal R}_n = 1 - \alpha_+ \, \zeta_+^n - \alpha_- \, \zeta_-^n  \, ,
\label{Rel2 de n double fan}
\end{equation}
with
\begin{eqnarray}
\zeta_{\pm} & = & \frac{1-p}{2} \, \left(1 + 2 \, p \, (1-p) \pm \sqrt{1+4 \, p^2 \, (1-p)^2} \right) \, , \label{zetaplusmoins double fan} \\
\alpha_{\pm} & = & \frac{1}{2} \pm \frac{1}{2} \, \frac{1+2 \, p^2}{\sqrt{1+4 \, p^2 \, (1-p)^2}} \, .
\label{alphaplusmoins double fan}
\end{eqnarray}
Here again --- if we forget that 1 is a third eigenvalue -- we have two eigenvalues $\zeta_{\pm}$. However, the situation is different from that of the $K_4$ ladder, because $\zeta_{\pm} \to 0$ when $p \to 1$, while $\zeta_+ \to 1$ and $\alpha_+ \to 1$ when $p \to 0$. We also expect that
\begin{equation}
{\mathcal R}_n \geq 1 - (1 - p^2)^n  \, ,
\label{Rel2 de n double fan minoree}
\end{equation}
the right-hand side of eq.~(\ref{Rel2 de n double fan minoree}) corresponding to $n$ elements of reliability $p^2$ in parallel.

%figure7
\begin{figure}[htb]
\centering
\includegraphics[scale=0.75]{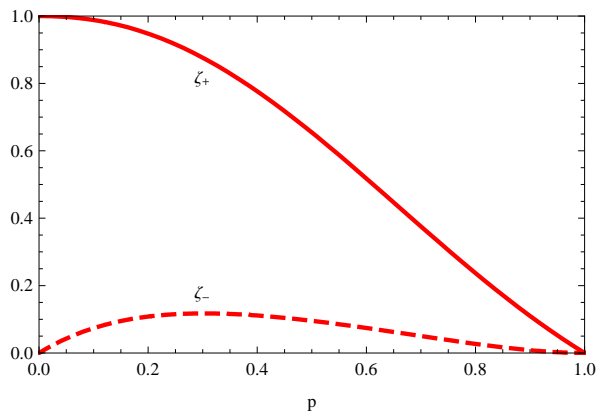}
\caption{$\zeta_{\pm}$ for the double fan.}
\label{zeta plus ou moins double fan}
\end{figure}

As $n$ increases, the contribution of $\zeta_+$ prevails over that of $\zeta_-$ (see Fig.~\ref{zeta plus ou moins double fan}), so that ${\mathcal R}_{n} \approx 1 - \alpha_+ \, \zeta_+^n$. Consequently,
\begin{equation}
\langle \, t \, \rangle_n \approx \frac{1}{\lambda} \, \int_0^1 \, \frac{dp}{p} \, \left(1 - \alpha_+ \, \zeta_+^n \right)\, .
\label{approximation Rn par zeta+ pour parallel-like}
\end{equation}
Because of the $1/p$ factor, the asymptotic expansion of $\langle \, t \, \rangle_n$ is now controlled by the behaviors of $\zeta_+$ and $\alpha_+$ for $p \to 0$:
\begin{eqnarray}
\zeta_{+} & = & 1 - p^2 - 2 \, p^3 + 2 \, p^4 + 4 \, p^5 - 8 \, p^6 - 4 \, p^7 + \cdots \, ,\\
\alpha_{+} & = & 1 + 2 \, p^3 - 8 \, p^5 + 12 \, p^6 + 24 \, p^7 + \cdots \, .
\end{eqnarray}
We can write
\begin{eqnarray}
1 - \alpha_+ \, \zeta_+^n & = & 1 - (1 - p^2)^n \nonumber \\
& & + (1-p^2)^n \, \left(1- \exp\left[- n \, (- \ln \zeta_+ + \ln (1-p^2) )\right] \right) \nonumber \\
& & + (1 - \alpha_+) \, \exp\left[- n \, (- \ln \zeta_+) \right]
\label{decomposition Rn double fan}
\end{eqnarray}
Each term on the right-hand side of eq.~(\ref{decomposition Rn double fan}) vanishes for $p \to 0$, so that the $1/p$ factor does not lead to a diverging integral in eq.~(\ref{approximation Rn par zeta+ pour parallel-like}). The first term of eq.~(\ref{decomposition Rn double fan}) gives the prevailing contribution, namely
\begin{eqnarray}
\frac{1}{\lambda} \, \int_0^1 \, \frac{dp}{p} \, \left( 1 - (1 - p^2)^n \right) & = & \frac{1}{2 \, \lambda} \, \int_0^1 \, \frac{dr}{r} \,  (1 - (1-r)^n) \nonumber \\
& = &  \frac{1}{2 \, \lambda} \, \int_0^1 \, \frac{ds}{1-s} \,  (1 - s^n) =  \frac{1}{2 \, \lambda} \, \sum_{i=0}^{n} \, \frac{1}{i} \, .
\label{decomposition Rn double fan terme 1}
\end{eqnarray}
This contribution is --- unsurprisingly --- half the usual result for $n$ elements in parallel, because the reliability $p^2$ translates into a $2 \, \lambda$ failure rate. For the two other contributions, the change of variable $\tau = n \, p^2$ gives a factor $\exp(-\tau)$; the remaining factors in eq.~(\ref{decomposition Rn double fan}) must be expanded in the vicinity of $\tau \to 0$, as in section~\ref{K_4 ladder}. Summing the three contributions gives
\begin{equation}
\lambda \, \langle \, t \, \rangle_n \rightarrow \frac{\ln n + {\mathbf C}}{2} + \frac{\sqrt{\pi}}{2 \, \sqrt{n}} - \frac{11}{4 \, n} + \frac{95 \, \sqrt{\pi}}{16 \, n^{3/2}} - \frac{1321}{24 \, n^2} + \cdots \, .
\label{MTTF asymptotique double fan}
\end{equation}

\begin{figure}[htb]
\centering
\includegraphics[scale=0.75]{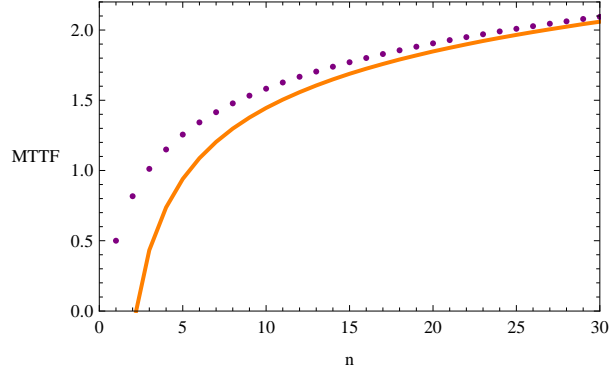}
\caption{Comparison between exact (dots) and asymptotic $\displaystyle \frac{\ln n + {\mathbf C}}{2} + \frac{\sqrt{\pi}}{2 \, \sqrt{n}} - \frac{11}{4 \, n}$ MTTF's for the double fan, in units of $\lambda^{-1}$.}
\label{Comparaison MTTF exact et asymptotique double fan}
\end{figure}

A comparison of the exact results with the asymptotic expansion in which we have kept the first three terms of eq.~(\ref{MTTF asymptotique double fan}) is plotted in Fig.~\ref{Comparaison MTTF exact et asymptotique double fan}. The agreement is good, even for moderate values of $n$.

%figure8

%%%%%%%%%%%%%%%%%%%%%%%%%%%%%%%%%%%%%
\subsection{Street $3 \times n$}
\label{Street 3xn}

%figure9
\begin{figure}[htb]
\centering
\includegraphics[scale=0.5]{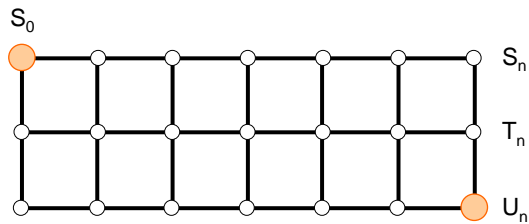}
\caption{Street $3 \times n$: the source is $S_0$, the destination $U_n$.}
\label{schema Street 3xn}
\end{figure}

In the preceding subsections, we have considered architectures for which ${\mathcal R}_n$ is exactly known through the analytic expressions of two eigenvalues $\zeta_{\pm}$. In complex systems, more than two eigenvalues may coexist, and be known only as roots of polynomial equations. However, the MTTF's asymptotic expression can still be derived from the knowledge of the generating function ${\mathcal G}(z)$ \cite{Stanley97} of the ${\mathcal R}_n$'s, namely
\begin{equation}
{\mathcal G}(z) = \sum_n \, {\mathcal R}_n \, z^n \, .
\label{definition generating function}
\end{equation}
Such is the case of the Street $3 \times n$, displayed in Fig.~\ref{schema Street 3xn}. This configuration has been studied for the two-terminal reliability ${\mathcal R}_n$ between $S_0$ and $U_n$ \cite{Tanguy06c,Theologou91,Carlier96,Kuo99,Yeh02,Yeh02conf,Rauzy03,Hardy07} (there is actually an offset of 1 between our $n$ and these references' $n$ because our source is $S_0$). For perfect nodes, ${\mathcal G}$ is given by ${\mathcal N}/({\mathcal D}_1 \, {\mathcal D}_2)$, with \cite{Tanguy06c}
\begin{eqnarray}
{\mathcal N} & = & p^2 - \left( 1 - p \right) \,p^4\,
   \left( 3 + 3\,p - 4\,p^2 \right) \,z \nonumber \\
   & & \hskip-1cm +
  {\left( 1 - p \right) }^3\,p^6\,
   \left( 2 + 11\,p - 3\,p^2 - 2\,p^3 \right) \,z^2 \nonumber \\
& & \hskip-1cm + {\left( 1 - p \right) }^3\,p^8\,
   \left( 2 - 4\,p + 3\,p^2 + 11\,p^3 - 13\,p^4 +
     3\,p^5 \right) \,z^3 \nonumber  \\
& & \hskip-1cm - {\left( 1 - p \right) }^4\,p^{10}\,
   \left( 3 + 6\,p - 12\,p^2 + 10\,p^3 - 10\,p^4 +
     4\,p^5 \right) \,z^4 \nonumber  \\
& & \hskip-1cm + {\left( 1 - p \right) }^6\,p^{12}\,
   \left( 1 + 8\,p - p^2 - 5\,p^3 - p^4 + p^5 \right)
     \,z^5 \nonumber  \\
& & \hskip-1cm - {\left( 1 - p \right) }^8\,p^{15}\,
   \left( 2 + 5\,p - 4\,p^2 \right) \,z^6 +
  {\left( 1 - p \right) }^{10}\,p^{18}\,z^7 , \\
{\mathcal D}_1 & = &  1 - \left( 1 - p^2 \right) \,p \,
   \left( 1 + p - p^2 \right) \,z \nonumber  \\
   & & \hskip-1cm +
  {\left( 1 - p \right) }^2\,p^3\,
   \left( 1 + p + p^2 - 2\,p^3 \right) \,z^2
   % \nonumber  \\ & &
   - {\left( 1 - p \right) }^4\,p^6\,z^3 ,
\\
{\mathcal D}_2 & = & 1 - p\,\left( 2 + 2\,p + p^2 -
9\,p^3 + 5\,p^4 \right)
     \,z \nonumber  \\
& &  + \left( 1 - p \right) \,p^2\,
   \left( 1 + 5\,p + 5\,p^2 - 6\,p^3 - 15\,p^4 \right. \nonumber \\
   & & \left. \hskip2.2cm +
     13\,p^5 + p^6 - 2\,p^7 \right) \,z^2 \nonumber  \\
& &  - {\left( 1 - p \right) }^2\,p^4\,
   \left( 2 + 6\,p + 6\,p^2 - 26\,p^3 + 17\,p^4 \right. \nonumber \\
   & & \left. \hskip2.2cm -
     18\,p^5 + 27\,p^6 - 16\,p^7 + 3\,p^8 \right) \,z^3 \nonumber  \\
& &  + {\left( 1 - p \right) }^4\,p^6\,
   \left( 1 + 6\,p + 4\,p^2 - p^3 - 17\,p^4 \right. \nonumber \\
   & & \left. \hskip2.2cm +
     9\,p^5 + 3\,p^6 - 2\,p^7 \right) \,z^4 \nonumber  \\
& &  -
  {\left( 1 - p \right) }^6\,p^9\,
   \left( 2 + 4\,p + p^2 - 7\,p^3 + 3\,p^4 \right) \,
   z^5 \nonumber  \\
& &  + {\left( 1 - p \right) }^8\,p^{12}\,z^6 .
\end{eqnarray}
${\mathcal N}$, ${\mathcal D}_1$ and ${\mathcal D}_2$ are polynomials in both $z$ and $p$.

The eigenvalue of greatest modulus, named $\zeta_+$ again, actually obeys ${\mathcal D}_2 = 0$ for $z = 1/\zeta_+$ (in the limit $p \to 1$, ${\mathcal D}_2 \to 1-z$ and $\zeta_+ \to 1$); all other eigenvalues tend to zero in that limit. Even though it is not possible to get an analytic expression for $\zeta_+$ as a function of $p$ (${\mathcal D}_2$ is of degree 6 in $z$), we can readily compute it numerically. We can also deduce from the constraint ${\mathcal D}_2(z = 1/\zeta_+) = 0$ the expansion of $\zeta_+$ as a function of $q$ for small $q$'s, starting with $\zeta_+ = 1$:
\begin{eqnarray}
\zeta_+ & = & 1 - q^3 - 4 \, q^4 - 4 \, q^5 + 14 \, q^6 + \cdots \, , \\
\label{zetaplus pres de 1}
- \ln \zeta_+ & = & q^3 + 4 \, q^4 + 4 \, q^5 - \frac{27}{2} \, q^6 + \cdots \, .
\label{- ln zetaplus pres de 1}
\end{eqnarray}
The scaling variable $\tau$ should be $n$ times the leading term of eq.~(\ref{- ln zetaplus pres de 1}), namely $\tau = n \, q^3$, so that
\begin{eqnarray}
\zeta_+^{n} & = & e^{- \tau} \, \exp - n \, \Big(4 \, \left( \frac{\tau}{n} \right)^{4/3} + 4 \, \left( \frac{\tau}{n} \right)^{5/3} + \cdots\Big) \nonumber \\
& = & e^{- \tau} \, \exp\Big( - 4 \, \tau^{4/3} \, n^{-1/3} - 4 \, \tau^{5/3} \, n^{-2/3} + \cdots \Big) \, .
\label{zetaplus puissance n 2}
\end{eqnarray}
$\alpha_+$ is deduced from $p$ and the numerical value of $\zeta_+$ through the residue of ${\mathcal G}$ at $z = 1/\zeta_+$. The general result is
\begin{equation}
\alpha_+ = \frac{- \zeta_+ \, {\mathcal N}\left(\frac{1}{\zeta_+}\right)}{{\mathcal D}'_z\left(\frac{1}{\zeta_+}\right)} \, ,
\end{equation}
where ${\mathcal D}$ is the denominator of ${\mathcal G}$ and $\displaystyle {\mathcal D}'_z = \frac{\partial {\mathcal D}}{\partial  z}$. Here, ${\mathcal D} = {\mathcal D}_1 \, {\mathcal D}_2$, leading to
\begin{equation}
\alpha_+ \to 1 - 2 \, q^2 - 4 \, q^3 + 7 \, q^4 + 22 \, q^5 + 20 \, q^6 + \cdots \, .
\label{alphaplus pres de 1}
\end{equation}

After inserting eqs.~(\ref{zetaplus puissance n 2}) and (\ref{alphaplus pres de 1}) in eq.~(\ref{approximation zeta+ puissance n en q 1}), and further series expansions and integration similar to those of Section~\ref{K_4 ladder}, the final asymptotic expansion reads
\begin{equation}
\lambda \, \langle \, t \, \rangle_n \rightarrow \frac{\Gamma(4/3)}{n^{1/3}} - \frac{5}{9} \, \frac{\Gamma(2/3)}{n^{2/3}} + \frac{7}{3 \, n} + \cdots.
\label{MTTF asymptotique street 3xn}
\end{equation}
Here again, it exhibits a power-law behavior, but $n^{-1/3}$ this time. In the next section, we relate the exponent to a specific property of the network.

%%%%%%%%%%%%%%%%%%%%%%%%%%%%%%%%%%%%%%%%%%%
%%%%%%%%%%%%%%%%%%%%%%%%%%%%%%%%%%%%%%%%%%%
\section{General case}
\label{General case}

In the preceding section, we have derived the asymptotic MTTF when the two-terminal reliability ${\mathcal R}_n$ is known, at least implicitly through a recursion relation. Here, we want to show that the leading terms of the MTTF and other moments may be obtained for arbitrary, recursive configurations. As shown in \cite{Tanguy06c}, ${\mathcal R}_n$ can be generally expressed as a product of transfer matrices --- whose size may be large but remains finite. For identical edge reliabilities $p$, the asymptotic behavior is controlled by the largest eigenvalue $\zeta_+$ of the (now unique) transfer matrix. We consider in the following architectures that look like some kind of ``series-like'' system, albeit more complex than those of the $K_4$ ladder and the Street $3 \times n$, or to a ``parallel-like'' one, like the double fan.

%%%%%%%%%%%%%%%%%%%%%%%%%%%%%%%%%%%%%%%%%%%
\subsection{``Series-like'' configuration}
\label{Series-like configuration}

This happens when the shortest path connecting the source to the destination has a length equivalent to $n$ as $n \to \infty$.
Here, we use again ${\mathcal R}_n \approx \alpha_+ \, \zeta_+^n$, and the MTTF is still controlled by the behavior of $\zeta_+$ and $\alpha_+$ in the vicinity of $p = 1$. The relevant expansions for $q \to 0$ have the form
\begin{eqnarray}
- \ln \zeta_+(1-q) & = & \alpha_{i} \, q^{i} + \alpha_{i+1} \, q^{i+1} + \cdots \, , \label{DL ln zetaplus general}\\
\alpha_+(1-q) & = & 1 + {\alpha'}_1 \, q + {\alpha'}_2 \, q^2 + \cdots \, ,
\label{DL alphaplus general}
\end{eqnarray}
from which
\begin{equation}
\langle \, t \, \rangle_n = \frac{1}{\lambda} \, \int_0^1 \, \frac{dq}{1-q} \, (1 + {\alpha'}_1 \, q + {\alpha'}_2 \, q^2 + \cdots) \, \exp\left[ - n \, (- \ln \zeta_+(1-q) - \alpha_{i} \, q^{i})\right] \, e^{- n \, \alpha_{i} \, q^{i}} \, .
\end{equation}
The adequate change of variable is now $\tau = n \, \alpha_i \, q^{i}$, or equivalently $\displaystyle q = \frac{\tau^{1/i}}{(n \, \alpha_i)^{1/i}}$; the upper bound of the integral, $n \, \alpha_i$, may again be replaced by $+ \infty$. The first term in the asymptotic expansion is therefore
\begin{eqnarray}
\langle \, t \, \rangle_n & \rightarrow & \frac{1}{\lambda} \, \int_0^{\infty} \, \frac{1}{i} \, \frac{1}{(n \, \alpha_i)^{1/i}} \, \tau^{1/i-1} \, d\tau \, e^{- \tau} \nonumber \\
& = & \frac{1}{\lambda} \, \frac{\Gamma(1+1/i)}{(n \, \alpha_i)^{1/i}} \, .
\end{eqnarray}
Likewise, in order to calculate the first term of the asymptotic expansion of $\lambda^2 \, \langle \, t^2 \, \rangle_n$, we have an additional factor $- 2 \, \ln (1-q)$, which is equivalent to $2 \, q$ when $q \to 0$ and brings an extra $\tau^{1/i}$. Finally,
\begin{equation}
\langle \, t^2 \, \rangle_n \rightarrow  \frac{1}{\lambda^2} \, \frac{\Gamma(1+2/i)}{(n \, \alpha_i)^{2/i}} \, ,
\end{equation}
so that the variance goes as
\begin{equation}
\langle \, t^2 \, \rangle_n - \langle \, t \, \rangle_n^2 \rightarrow  \, \frac{1}{\lambda^2} \, \frac{\Gamma(1+2/i) - \Gamma(1+1/i)^2}{(n \, \alpha_i)^{2/i}} \, ,
\end{equation}
from which
\begin{equation}
\frac{\sqrt{\langle \, t^2 \, \rangle_n - \langle \, t \, \rangle_n^2}}{\langle \, t \, \rangle_n} \rightarrow  \, \sqrt{\frac{\Gamma(1+2/i)}{\Gamma(1+1/i)^2}-1} \, ,
\end{equation}
which is independent of $n$. Actually, it only depends on $i$, which is the lowest order of the $q$-dependence of $1-\zeta_+$ when $q \to 0$ (see eq.~(\ref{DL ln zetaplus general})).

For the higher moments $\langle \, t^m \, \rangle_n$, the generalization is straightforward; we can also go beyond the first order in the expansion, following the recipe of the preceding section. Setting $\eta = (n \, \alpha_i)^{-1/i}$, we find
\begin{eqnarray}
\lambda^m \, \langle t^m \rangle_n & = & \eta^m \, \Bigg\{ \hskip7mm \Gamma\left(1+\frac{m}{i}\right) \nonumber \\
& & \hskip1cm + \eta \, \frac{m}{i} \, \Bigg[ \frac{1}{2} \, (1+m+2 \, \alpha'_1) \, \Gamma\left(\frac{1+m}{i}\right) - \frac{\alpha_{i+1}}{\alpha_{i}} \,  \Gamma\left(1+\frac{1+m}{i}\right) \Bigg] \nonumber \\
& & \hskip1cm + \eta^2 \, \frac{m}{i} \, \Bigg[ \frac{10+11 \, m +3 \, m^2 + 12 \, (1+m) \, \alpha'_1 + 24 \, \alpha'_2}{24} \, \Gamma\left(\frac{2+m}{i}\right) \nonumber \\
& & \hskip2.6cm - \left( \frac{\alpha_{i+2}}{\alpha_{i}} + \frac{\alpha_{i+1}}{\alpha_{i}} \, \frac{1+m + 2 \, \alpha'_1}{2} \right) \, \Gamma\left(1+\frac{2+m}{i}\right) \nonumber \\
& & \hskip26mm + \frac{1}{2} \, \left( \frac{\alpha_{i+1}}{\alpha_{i}} \right)^2 \, \Gamma\left(2+\frac{2+m}{i}\right) \Bigg] + \cdots \Bigg\} \, .
\label{resultat final expansion moment ordre m}
\end{eqnarray}

\begin{figure}[htb]
\centering
\includegraphics[scale=0.75]{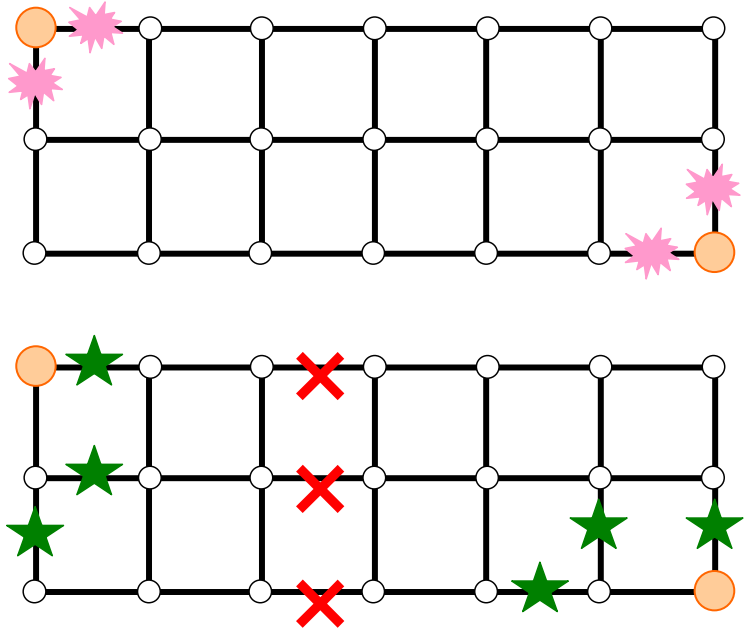}
\caption{Cuts of order 2 (top) and 3 (bottom) for the Street $3 \times n$. Failed edges are indicated by stars and crosses.}
\label{coupures d'ordre 2 et 3 pour les Street 3xn}
\end{figure}

We may wonder: is this result merely formal, or is it actually possible to determine $\lambda^m \, \langle t^m \rangle_n$ for an arbitrary, recursive network ? The answer to this question is yes. Even though we do not know the exact value of ${\mathcal R}_n$ or the associated greatest eigenvalue $\zeta_+$, we can still infer the $\alpha_k$'s and $\alpha'_j$'s appearing in eqs.~(\ref{DL ln zetaplus general})--(\ref{DL alphaplus general}) because for large $n$, ${\mathcal R}_n \approx \alpha_+ \, \zeta_+^n$. These parameters can be deduced from the expansion of the unavailability ${\mathcal U}_n = 1-{\mathcal R}_n$ for $q \to 0$. We have
\begin{eqnarray}
{\mathcal U}_n & \approx & 1 - \alpha_+ \, e^{- n (-\ln \zeta_+)} \nonumber \\
& = & 1 - (1 + {\alpha'}_1 \, q + {\alpha'}_2 \, q^2 + \cdots) \, \exp \left[ - n \, \left(\alpha_{i} \, q^{i} + \alpha_{i+1} \, q^{i+1} + \cdots \right) \right]
\,
\label{approximation unavailability}
\end{eqnarray}
and must keep track of the successive powers of $q$, along with their dependence with $n$. Let us illustrate this claim with the Street $3 \times n$ case. A simple cut enumeration gives (when $n$ is large, so as to avoid ``boundary'' effects)
\begin{equation}
{\mathcal U}_n = 2 \, q^2 + (n + 4) \, q^3 + (4 \, n - 7) \, q^4 + \cdots \, .
\label{unavailability street 3xn enumeration of cuts}
\end{equation}
The first term of the right-hand side of eq.~(\ref{unavailability street 3xn enumeration of cuts}) is easy to obtain. There are only two cuts of order 2 preventing a connection between source and destination (see the top of Fig.~\ref{coupures d'ordre 2 et 3 pour les Street 3xn}), hence the $2 \, q^2$ term. For the cuts of order 3, different possibilities occur as displayed at the bottom of Fig.~\ref{coupures d'ordre 2 et 3 pour les Street 3xn}. Firstly, three parallel links may fail; there are $n$ such instances. Secondly, close to the source or the destination, there are four triple failures (only two are represented by green stars in Fig.~\ref{coupures d'ordre 2 et 3 pour les Street 3xn}, the remaining ones can be deduced by symmetry). This gives the $(n + 4) \, q^3$ term. A comparison between eqs.~(\ref{approximation unavailability}) and (\ref{unavailability street 3xn enumeration of cuts}) gives ${\alpha'}_1 = 0$, ${\alpha'}_2 = -2$, ${\alpha'}_3 = -4$, and ${\alpha}_3 = 1$. From these values, we get
\begin{equation}
\lambda^m \, \langle t^m \rangle_n = \frac{1}{n^{1/3}} \, \left( \Gamma\left(1+\frac{m}{3}\right) - \frac{5}{6} \, \frac{1}{n^{1/3}} \, \Gamma\left(1+\frac{m+1}{3}\right) + \cdots \right) \, ,
\end{equation}
which agrees with eq.~(\ref{MTTF asymptotique street 3xn}) when $m = 1$.

In conclusion, the exponent of the power-law behavior in $n$ is nothing but the inverse of the number of necessary cuts to isolate each elementary cell from its neighbors. Note that $\alpha_i$ is not necessarily equal to 1, since it represents the number of independent cuts of order $i$.

%figure10

%%%%%%%%%%%%%%%%%%%%%%%%%%%%%%%%%%%%
\subsection{``Parallel-like'' configuration}
\label{Parallel-like configuration}

We can also consider configurations where the reliability is asymptotically equal to 1 when $n \to \infty$, as in section~\ref{Double fan}. In the vicinity of $p \to 0$ (the relevant domain here), ${\mathcal R}_n \approx 1 - \alpha_+ \, \zeta_+^n$ and the needed expansions are
\begin{eqnarray}
-\ln \zeta_+ & = & \beta_{i} \, p^{i} + \beta_{i+1} \, p^{i+1} + \cdots \, ,\label{ln zetaplus p tend vers 0}\\
\alpha_+ & = & 1 + {\beta'}_1 \, p + {\beta'}_2 \, p^2 + \cdots \,  , \label{alphaplus p tend vers 0}
\label{DL pour approximation availability}
\end{eqnarray}
so that
\begin{equation}
1 - \alpha_+ \, \zeta_+^n = \underbrace{\left(1 - (1-p^{i})^{\beta_i \, n}\right)}_{\rm A} + \underbrace{\left((1-p^{i})^{\beta_i \, n} - \zeta_+^n\right)}_{\rm B} + \underbrace{(1 - \alpha_+) \, \zeta_+^n}_{\rm C} \, .
\label{Decomposition en A B C}
\end{equation}
The contribution of A is (omitting the $\lambda^{-1}$ factor)
\begin{equation}
{\rm A} \rightarrow \int_0^1 \, \frac{dp}{p} \, \left( 1 - (1-p^{i})^{\beta_i \, n} \right) = \frac{1}{i} \, \sum_{k = 1}^{\beta_i \, n} \, \frac{1}{k} \rightarrow \frac{1}{i} \, \left( \ln \beta_i \, n + {\mathbf C} + \frac{1}{2 \, \beta_i \, n} + \cdots \right)\, ,
\label{valeur de A}
\end{equation}
The contribution of B depends on the value of $i$ because
\begin{equation}
-\ln \zeta_+ + \beta_{i} \, \ln(1 - p^{i}) = \beta_{i+1} \, p^{i+1} + \cdots \; \; + \beta_{i} \left(- \frac{1}{2} \, p^{2 \, i} + \cdots \right)\, .
\end{equation}
If $i = 1$, we must take the two terms of degree 2 into account; otherwise, only the $\beta_{i+1} \, p^{i+1}$ term needs be kept. After asymptotic expansions similar to those performed in the preceding sections, we have
\begin{eqnarray}
{\rm B} & \rightarrow & \left(\beta_2 - \frac{\beta_1}{2} \right) \, \frac{1}{\beta_1^2} \, \frac{1}{n} + \cdots \hskip25mm (i = 1) \\
& \rightarrow & \frac{\beta_{i+1}}{i} \, \Gamma\left(1+ \frac{1}{i}\right) \, \frac{1}{\beta_i^{1+1/i}} \, n^{-1/i} + \cdots \hskip5mm (i > 1) \, .
\label{valeurs de B1 et B2}
\end{eqnarray}
The contribution of C is easier to compute because $1 - \alpha_+$ vanishes as $p \to 0$, thereby compensating the singular term $1/p$ in the integral. With the change of variable  $\tau = n \, \beta_i \, p^{i}$, we get
\begin{equation}
{\rm C} \rightarrow \frac{-\beta'_1}{(n \, \beta_{i})^{1/i}} \, \Gamma\left(1+ \frac{1}{i}\right) + \cdots\, .
\label{valeur de C}
\end{equation}
Note that the $n$-dependence of C is $n^{-1/i}$, which decreases less rapidly than $1/n$ if $i > 1$.
The sum of contributions A, B, and C finally expands as (for $i \geq 1$)
\begin{equation}
\lambda \, {\rm MTTF}_n \rightarrow \frac{1}{i} \, \left( \ln (\beta_i \, n) + {\mathbf C} \right) + \left( \frac{\beta_{i+1}}{i \, \beta_{i}} - \beta'_{1} \right) \, \frac{\Gamma\left(1+\frac{1}{i}\right)}{(n \, \beta_{i})^{1/i}} + \cdots
\label{valeur de A plus B plus C}
\end{equation}

As in the preceding subsection \ref{Series-like configuration}, the coefficients $\beta_{j}$, $\beta'_{k}$, etc. may be deduced by evaluating the availability through a path enumeration in the limit $p \to 0$. For instance, in the case of the double fan
\begin{equation}
{\mathcal R}_n = n \, p^2 + 2 \, (n-1) \, p^3 + \cdots
\label{debut Rn pour double fan p tend vers 0}
\end{equation}
This must be compatible with the expansions of $\alpha_+$ and $- \ln \zeta_+$ in eqs.~(\ref{ln zetaplus p tend vers 0})--(\ref{alphaplus p tend vers 0}). Because eq.~(\ref{debut Rn pour double fan p tend vers 0}) has no linear term, $\beta_1 = \beta'_1 = 0$. The coefficient of $p^2$ being equal to $n$, we deduce $i = 2$, $\beta_2 = 1$, and $\beta'_2 = 0$; the coefficient of $p^3$ then implies $\beta_3 = \beta'_3 = 2$. Inserting these values in eq.~(\ref{valeur de A plus B plus C}) gives back the first two terms of eq.~(\ref{MTTF asymptotique double fan}).

%%%%%%%%%%%%%%%%%%%%%%%%%%%%%%%%%%%%%%%%%%%%%%%%%%%%%%%%%%%%
\section{Approximate reliability of large, recursive, ``series-like'' systems}
\label{Approximate reliability of large recursive systems}

We have seen that the asymptotic expansion of the MTTF and the higher moments for a large, recursive system can be obtained with minimal effort. In the same line of thought, is it possible to find an approximate reliability such that all its  moments give the same value as the true reliability, at least for the first terms of the expansion in $n$. The first term of eq.~(\ref{resultat final expansion moment ordre m}), namely
\begin{equation}
\langle t^m \rangle_n = \frac{\Gamma(1+m/i)}{\lambda^m \, (n \, \alpha_i)^{m/i}} \, ,
\label{premier terme moment ordre m}
\end{equation}
reminds us of what would be obtained for a Weibull distribution \cite{Shooman68,KuoZuo}. Indeed, it is straightforward to show that
\begin{equation}
{R}^{(0)}_n(t) = \exp\left( - n \, \alpha_i \, \lambda ^{i} \, t^{i} \right) \, ,
\label{premiere dispo equivalente}
\end{equation}
would give eq.~(\ref{premier terme moment ordre m}) exactly.

Is it possible to improve this expression, i.e., propose an effective reliability leading to the correct first two terms in the asymptotic expansion of each moment $\langle t^m \rangle_n$ ? Provided that ${\alpha'}_1 = 0$, i.e., that there is no cut of order one (it would then be easy to factor out this link contribution, and proceed with the remaining parts of the system), the answer is again positive.

\begin{figure}[htb]
\centering
\includegraphics[scale=0.75]{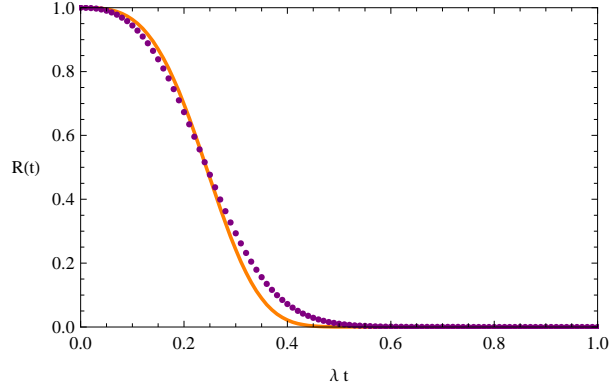}
\caption{Exact (dots) and asymptotic (full line) reliability for the Street $3 \times n$ architecture ($n = 30$).}
\label{Comparaison fiabilites vraie et asymptotique street 3xn et n egal 30}
\end{figure}

Calculating the moments of
\begin{equation}
{R}^{(1)}_n(t) = \exp\left[ - n \, (\alpha_i \, \lambda ^{i} \, t^{i} + \widetilde{\alpha}_{i+1} \, \lambda ^{i+1} \, t^{i+1}) \right] \, ,
\label{deuxieme dispo equivalente}
\end{equation}
we find that
\begin{equation}
\lambda^m \, \langle t^m \rangle_n = \frac{1}{(n \, \alpha_i)^{m/i}} \, \left( \Gamma\left(1+\frac{m}{i}\right) + \frac{\widetilde{\alpha}_{i+1}}{\alpha_i} \, \frac{1}{(n \, \alpha_i)^{m/i}} \, \left( - \frac{m}{i} \, \Gamma\left(1+\frac{m+1}{i}\right) \right) + \cdots \right) \, .
\label{moments fiabilite1}
\end{equation}
Equations~(\ref{resultat final expansion moment ordre m}) and (\ref{moments fiabilite1}) match if
\begin{equation}
\frac{\widetilde{\alpha}_{i+1}}{\alpha_i} = \frac{\alpha_{i+1}}{\alpha_i} - \frac{1}{2} \, \frac{m + 1 + 2 \, {\alpha'}_1}{\frac{m + 1}{i}} \, ,
%\label{}
\end{equation}
so that for ${\alpha'}_1 = 0$, the constraint is satisfied when
\begin{equation}
\widetilde{\alpha}_{i+1} = \alpha_{i+1} - \frac{i}{2} \, {\alpha_i} \, .
\label{condition sur alpha tilde i+1}
\end{equation}
This finally gives
\begin{equation}
{R}^{(1)}_n(t) = \exp \left[- n \, \left(\alpha_i \, \lambda ^{i} \, t^{i} + \left(\alpha_{i+1} - \frac{i}{2} \, {\alpha_i}\right) \, \lambda ^{i+1} \, t^{i+1} \right) \right] \, .
\label{deuxieme dispo equivalente finale}
\end{equation}
This expression slightly improves over the Weibull distribution of eq.~(\ref{premiere dispo equivalente}). In the case of the Street $3 \times n$,
\begin{equation}
{R}^{(1)}_n(t) = \exp\left[ - n (\lambda^3 \, t^3 + \frac{5}{2} \, \lambda^4 \, t^4)\right] \, .
\label{dispo equivalente finale street 3xn}
\end{equation}
This expression and the true reliability are plotted in Fig.~\ref{Comparaison fiabilites vraie et asymptotique street 3xn et n egal 30}; the agreement is already satisfying for $n = 30$.

%figure11

%%%%%%%%%%%%%%%%%%%%%%%%%%%%%%%%%%%%%%%%%%%%%%%%%%%%%%%%%%%%
\section{Non-exponential distribution functions}
\label{Nonexponential distribution functions}

In the preceding sections, we have considered elements whose reliability is $p(t) = \exp(- \lambda \, t)$. Although this distribution is often chosen because calculations are simpler, other models may be used: Weibull, gamma, lognormal, etc. \cite{Shooman68,KuoZuo}. We investigate here the influence of the true $p(t)$ on the $n$-dependence of the MTTF and higher moments for ``series-like'' systems.

We can invert $p(t)$ as $t = \chi(p)$, so that eq.~(\ref{integrale de Rel sur p}) transforms into
\begin{equation}
{\rm MTTF}_n = \langle \, t \, \rangle_n = \int_0^1 \, dp \, {\mathcal R}_n(p) \, (- \chi'(p)) \, .
\label{MTTF avec chi(p)}
\end{equation}
For higher moments, we would get
\begin{equation}
\langle \, t^m \, \rangle_n = m \, \int_0^1 \, dp \, \chi^{m-1}(p) \, {\mathcal R}_n(p) \, (- \chi'(p)) \, .
\label{mu_m avec chi(p)}
\end{equation}

For ``series-like'' configurations, we have again to consider what happens for $p \to 1$, or equivalently for $t \to 0$. Assuming that asymptotically
\begin{equation}
- \chi'(p) \rightarrow a_{\beta} \, (1-p)^{\beta} \, ,
\label{limite chi'(p) pour p tend vers 1}
\end{equation}
we have (because $\chi(1) = 0$)
\begin{equation}
\chi(p) = t \rightarrow \frac{a_{\beta}}{\beta+1} \, (1-p)^{\beta+1} \, ,
\label{limite chi(p) pour p tend vers 1}
\end{equation}
so that
\begin{equation}
p \rightarrow 1 - \left( \frac{(\beta+1) \, t}{a_{\beta}} \right)^{1/(\beta+1)} \, .
\label{limite chi'(p) pour p tend vers 1}
\end{equation}

Keeping $- \ln \zeta_+ \to \alpha_i \, q^{i}$ and $\alpha_+ \to 1$, we get to lowest order
\begin{eqnarray}
\langle \, t^m \, \rangle_n & \rightarrow & m \, \int_0^1 \, dq \, \left( \frac{a_{\beta}}{\beta+1} \right)^{m-1} \, q^{(\beta+1) \, (m-1)} \, a_{\beta} \, q^{\beta} \, \exp\left( - n \, \alpha_i \, q^{i}\right) \nonumber \\
& \rightarrow & \left( \frac{a_{\beta}}{\beta+1} \right)^{m} \, \frac{1}{(n \, \alpha_i)^{(\beta+1) \, m/i}} \, \Gamma\left(1 + \frac{(\beta+1) \, m}{i}\right)\, .
\label{t puissance m asymptotique}
\end{eqnarray}
The $n$-dependence of $\langle \, t^m \, \rangle_n$ is therefore affected by the structure of the graph (through $i$ and $\alpha_i$) and by the true failure-time distribution of each element (through $a_{\beta}$ and $\beta$). The exponent of the power-law asymptotic behavior is $\displaystyle \frac{(\beta+1) \, m}{i}$, and the total reliability goes asymptotically as
\begin{equation}
{R}_n(t) \approx \exp\left[ - n \, \alpha_i \, \left( \frac{(\beta+1) \, t}{a_{\beta}} \right)^{1/(\beta+1)} \right] \, .
\label{effective Rn avec distribution non exponentielle}
\end{equation}

\section{Conclusion}

We have shown that very simple asymptotic expansions may be obtained for the mean time to failure (and higher moments) for general recursive, meshed networks. By contrast with the simple series and parallel systems considered in many textbooks, the size-dependence of the MTTF of a ``series-like'' system follows a power-law behavior, whose exponent is linked to the number of cuts disconnecting an elementary cell to its neighbors. Comparison with the exact results for various architectures show that the agreement is often reached when the system contains a few dozens of the (repeated) pattern structure. A simple, approximate expression for the effective global reliability of the system has also been proposed, which is very simple to derive by a mere enumeration of cut-sets or path-sets.

The calculations have been performed in the context of the two-terminal reliability of general systems; they obviously apply to all-terminal reliability, and would belong to the ``series-like'' category.

\section*{Acknowledgment}
Useful and stimulating discussions with Nancy Perrot, Guillaume Boulmier, Matthieu Chardy, Bertrand Decocq, S\'{e}bastien Nicaisse, and Mathieu Trampont are gratefully acknowledged.

\end{document}